# 3D printed microchannels for sub-nL NMR spectroscopy

E. Montinaro[a], M. Grisi[a], M. C. Letizia[a], L. Pethö[b], M. A. M. Gijs[a], R. Guidetti[c], J. Michler[b], J. Brugger[a], and G. Boero[a,*]

[a]Ecole Polytechnique Fédérale de Lausanne (EPFL), Laboratory for Microsystems, 1015, Lausanne, Switzerland
[b]Swiss Federal Laboratories for Materials Science and Technology (EMPA), Laboratory for Mechanics of Materials and Nanostructures, 3602, Thun, Switzerland
[c]University of Modena and Reggio Emilia, Department of Life Sciences, 41125, Modena, Italy

Nuclear magnetic resonance (NMR) experiments on subnanoliter (sub-nL) volumes are hindered by the limited sensitivity of the detector and the difficulties in positioning and holding such small samples in proximity of the detector. While the sensitivity of inductive detectors is optimized by miniaturizing the dimensions of the inductor to match the size of the specimen, the sample handling becomes more difficult as the size of the detector is scaled down. In this work, we report on NMR experiments on liquid and biological entities immersed in liquids having volumes down to 100 pL. These measurements are enabled by the fabrication of high spatial resolution 3D printed microfluidic structures, specifically conceived to guide and confine sub-nL samples in the sub-nL most sensitive volume of a single-chip integrated NMR probe. The microfluidic structures are fabricated using a two-photon polymerization 3D printing technique having a resolution better than 1 µm$^3$. The NMR probe consists of an electronic transceiver and a 150 µm diameter excitation/detection microcoil, co-integrated on a single silicon chip of about 1 mm$^2$. The high spatial resolution 3D printing approach adopted here allows to rapidly fabricate complex microfluidic structures tailored to position, hold, and feed biological samples, with a design that maximizes the NMR signals amplitude and minimizes the static magnetic field inhomogeneities. The layer separating the sample from the microcoil, crucial to exploit the volume of maximum sensitivity of the detector, has a thickness of 10 µm. To demonstrate the potential of this approach, we report NMR experiments on sub-nL intact biological entities in liquid media, specifically ova of the tardigrade *Richtersius coronifer* and sections of *Caenorhabditis elegans* nematodes. We show a sensitivity of 2.5x10$^{13}$ spins/Hz$^{1/2}$ on $^1$H nuclei at 7 T, sufficient to detect highly concentrated endogenous compounds in active volumes down to 100 pL in a measurement time of 3 hours. Spectral resolutions of 0.01 ppm in liquid samples and of 0.1 ppm in the investigated biological entities are also demonstrated. The obtained results indicate a promising route for NMR studies at the single unit level of important biological entities having sub-nL volumes, such as living microscopic organisms and eggs of several mammalians, humans included.

## Introduction

Methods based on the nuclear magnetic resonance (NMR) phenomenon are widely used in physics, chemistry, medicine and biology.[1-3] During an NMR experiment, the sample is placed in a static magnetic field and excited with electromagnetic fields at frequencies and strengths which have no biological effects. Due to this property and to its resolving power, NMR is successfully applied, e.g., to diagnostic imaging[4,5] and in-vivo spectroscopy[6-8] of large living animals. The use of NMR methodologies for the study of sub-µL volumes is hindered by sensitivity limitations. The search for methods enabling the extensions of this powerful technique to the study of smaller volumes is an active research domain. These efforts include the miniaturization of inductive methods[9-43] as well as the use of more sensitive but less versatile non-inductive approaches.[44-48] Various techniques permitted the optimization of NMR inductive detectors for volume ranging from 1 µL down to a few nL. Some of these were used to perform pioneering studies of small collections of microorganisms,[22, 49-51] perfused tumor spheroids,[34] and large single cells and embryos.[52-58] NMR-based studies of intact single biological entities were, until recently, demonstrated down to volumes of 5 nL[51, 54, 55] whereas typical volumes of most cells and microorganisms are below the nL scale.[59] Recently, we reported the use of ultra-compact single-chip NMR probes as convenient tools to deliver state-of-art spin sensitivity for sub-nL volumes.[33, 41] Such probes, entirely realized on a single 1 mm$^2$ complementary-metal-oxide-semiconductor (CMOS) microchip, consist of a multilayer microcoil and a co-integrated low-noise electronic transceiver. Thanks to the achieved sensitivity, we were able to perform the first NMR spectroscopy studies of single untouched sub-nL ova of microorganisms having active volumes down to 0.1 nL.[41] These experiments were conducted by manually placing the sample on top of the microcoil using a polystyrene cup filled with agarose gel.[41] This study demonstrated the detection of highly concentrated endogenous compounds, but improvements concerning sample manipulation are required to enable long-lasting experiments in more biocompatible environments. In particular, the study of biological samples in liquids would facilitate a non-invasive sample handling and it would provide a more biocompatible and controllable environment during the experiments. To perform NMR spectroscopy studies on small biological entities in a liquid environment, relatively complex microfabrication techniques are required, in particular for the placing and holding of the samples in the sensitive region of the miniaturized detector. Several approaches to combine microfluidic structures with microsolenoids[10, 22, [29, 32, 60, 61] and planar microcoils[16, 60, 62] have been reported to date. Various techniques were developed to shrink the size of solenoids and pattern them on capillaries[43, 63-65] or around hollow pillars via wire bonding.[29, 40] At the sub-nL volume scales, an efficient combination of microsolenoids with microfluidic structures would requires significantly more complex microfabrication processes than the one for planar microcoils with planar microfluidic structures.

In this work, we report on the design and fabrication of high spatial resolution microfluidic structures capable to position and hold sub-nL biological entities in the sub-nL sensitive volume of a planar microcoil, used for NMR signal excitation/detection, co-integrated on the same silicon chip with the transceiver electronics (Fig. 1a). The single-chip CMOS detector offer a robust planar working surface, but its very small detection volume (about 0.2 nL) sets challenging fabrication constraints for the microfluidic design, which must hold the sample in close proximity of the microcoil without introducing significant static magnetic field inhomogeneities. To overcome these challenges, we fabricated microfluidic structures using a high spatial resolution 3D printer (Photonic Professionals GT, Nanoscribe GmbH, Germany), based on a two-photon polymerization process [66-68] and having a resolution better than 1 µm³. With this approach we managed to fabricate microchannels capable to confine the samples under investigation at distances of about 10 µm from the microcoil surface. The proximity between the microcoil and the sample confined in the microfluidic channel is crucial to preserve a high effective spin sensitivity. To demonstrate the validity of our approach, we report NMR measurements on sub-nL samples (and sub-nL portion of larger samples) having significantly different size, geometry, and nature. We report experiments on liquids, where we show spectral resolutions down to 0.007 ppm full width at half maximum (i.e., about 2 Hz at 300 MHz) in liquid samples of 100 pL. Additionally, we report experiments on two radically different biological entities, i.e. tardigrade *Richtersius coronifer* ova and nematode *Caenorhabditis elegans* worms. Despite the tiny size (about 100 pL) and the broad intrinsic linewidth (about 30 Hz at 300 MHz) of these samples, the achieved sensitivity ($2.5 \times 10^{13}$ spins/Hz$^{1/2}$) is sufficient to detect highly concentrated endogenous compounds. The microfluidic channels are connected to a robust fluidic interface that tolerates the application of flows as strong as 7 µL/s and guarantees an efficient sealing for several days (experiments with a duration of two weeks have been successfully performed). The results of the reported experiments indicates that the approach proposed here allows for the non-invasive and efficient handling and trapping of living entities for NMR investigations at the sub-nL volume scale, in conditions of high sensitivity and sample limited spectral resolution.

## Materials and methods

### Sensitive volume of the microcoil and related microfabrication constraints

The single chip NMR detector, described in details in Ref. 33, has an integrated excitation/detection multilayer microcoil having a diameter of 150 µm (Fig. 1a). The planar geometry and the localized sensitive volume (the distance between the microcoil and the co-integrated RF preamplifier is less than 300 µm) allow for an eased approach and assembly with microfluidic structures. The integrated planar microcoil has a high sensitivity in a distorted spherical volume having a diameter of about 100 µm. In the following, we quantify the dependence of the local sensitivity as a function of the distance from the microcoil center. As a convention throughout the article, the planar microcoil lies in the $yz$ plane, with the static magnetic field $B_0$ along the $z$ axis. Using the principle of reciprocity,[69] the signal contribution per elementary volume $dS(\mathbf{r})$ (i.e., the local sensitivity) is proportional to $B_{uxy}(\mathbf{r})\sin(\theta)$, where $B_{uxy}(\mathbf{r}) = ((B_{ux}(\mathbf{r}))^2+B_{uy}((\mathbf{r}))^2)^{1/2}$ is the field component in the $xy$ plane (i.e., perpendicular to $B_0$) produced by the microcoil carrying a unitary current, $\theta = \gamma B_{uxy}(\mathbf{r})I\tau$ is the flip angle, $\gamma$ is the nuclear gyromagnetic ratio, $\tau$ is the pulse length, and $I$ is the current carried by the coil during the excitation pulse.[1] The maps of sensitivity of the integrated microcoil shown in Fig. 1b are obtained computing $B_{uxy}(\mathbf{r})\sin(\theta)$, starting from the computation of $B_{uxy}(\mathbf{r})$ via a Biot-Savart code implemented in Matlab. As shown in Fig. 1b, the sensitivity of the microcoil decreases rapidly with the distance along the $x$ axis. As a result, the microfluidic system used to hold the sample on top of the microcoil should have a thin separation layer. Variation of the pulse length $\tau$ implies a variation of the local flip angle $\theta$, thus a variation of the spatial distribution of the sensitivity. As shown in Fig. 2 and described in details in the next subsections, we fabricated two microchannels having different height and width. The integrated NMR signal $S$ is proportional to the integral of $dS(\mathbf{r})$ over the total volume of the microchannel. Figure 1b shows the sensitivity maps for $\tau = 3.7$ µs, which is the pulse length that maximizes the integrated signal for both microchannels, assuming a separation $d$ between the microcoil and the sample of 10 µm. Upon normalization of the integrated NMR signal to $S_{(d=0\,\mu m)}=1$ for the case of the channel volume placed in direct contact with the chip, the signals $S_{(d=10\,\mu m)}$ and $S_{(d=20\,\mu m)}$, evaluated in conditions of optimal $\tau$, are respectively of about 0.7 and 0.5 for both microchannels. Hence, a separation layer of 10 µm (Figs. 1c and 1d) is an acceptable compromise between the loss of sensitivity and the robustness of the structure.

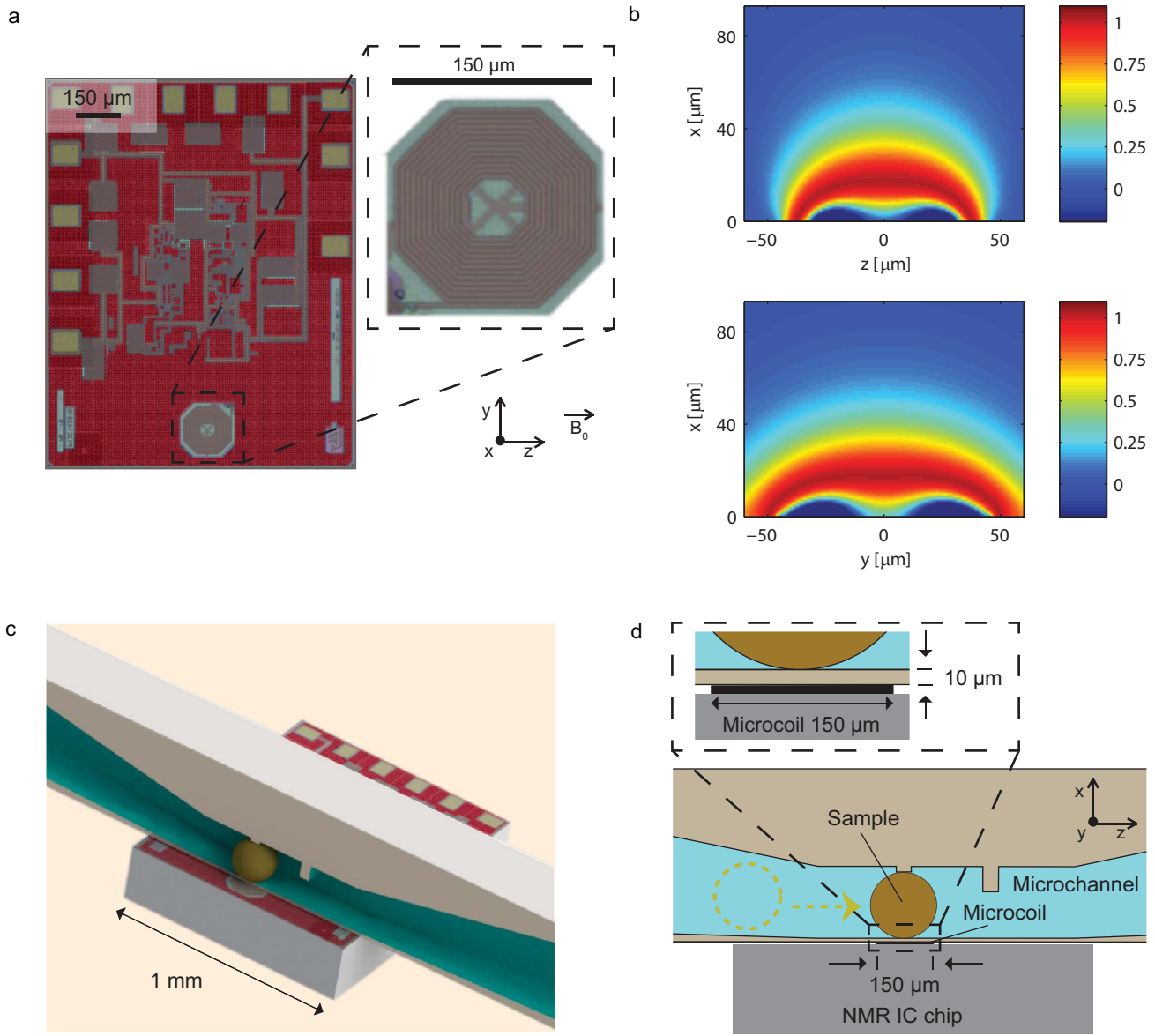

**Fig. 1**: (**a**) Microphotograph of the NMR single-chip detector, fabricated using a 130 nm CMOS technology from STMicroelectronics. The integrated planar microcoil is realized using four copper metal layers. The total number of turns is 22 and the outer diameter is 150 μm (see inset). The details of the microcoil and of the co-integrated electronics are reported in Ref. 33. (**b**) Sensitivity maps of the microcoil (the sensitivity is defined as $B_{uxy}(\mathbf{r})\sin(\theta)$, see main text). The microcoil lies in the *yz* plane and the static magnetic field $B_0$ lies along the *z* axis. The sensitivity in the *xz* (top) and *xy* (bottom) planes are computed for a pulse length $\tau$ = 3.7 μs and an excitation current *I* = 9 mA. Dark red color identifies the region of maximum sensitivity. (**c**) Rendered image of the single chip integrated detector combined with the 3D printed microfluidic structure. (**d**) Illustration of the approach used to place the sample onto the most sensitive area of the excitation/detection microcoil. The flow drives the sample in proximity of the integrated microcoil. The dashed arrow indicates the direction of insertion of the ovum, which is trapped by two 10-μm-high pillars. A 50-μm-high pillar is employed to block the sample in case of accidental overpressure. A 10-μm-thick crosslinked IP-S photoresist layer defines the spacing between the sample and the chip surface (see inset).

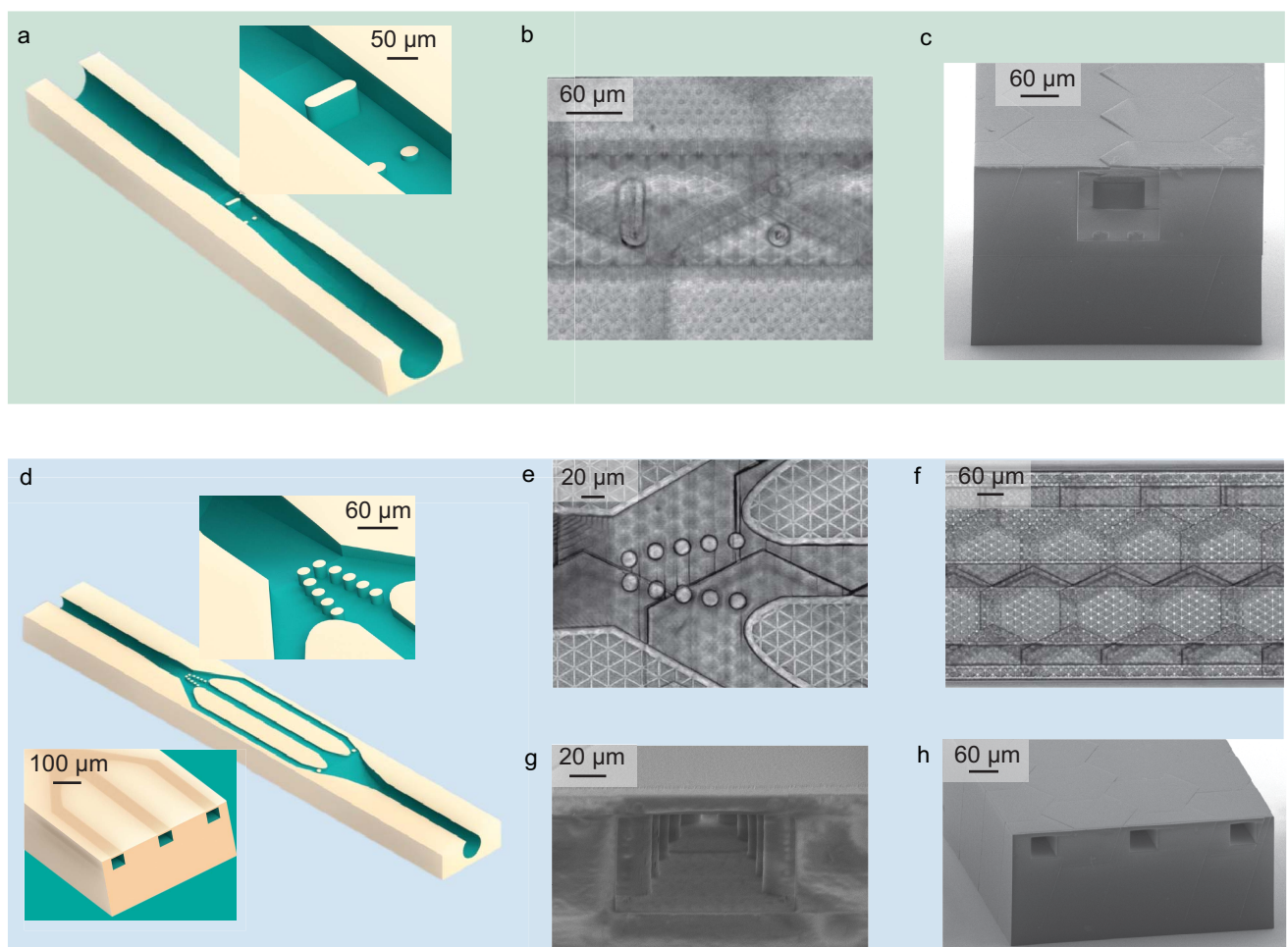

**Fig. 2**: (**a**) Schematic of the microchannel used to trap a single *Rc* ovum. A close up view of the trapping pillars is shown. Two 10-μm-high pillars trap the ovum, while a second pillar is implemented to prevent the loss of the ovum in case of accidental overpressure. (**b**) Top view optical microscope picture of the central part of the microchannel. (**c**) SEM picture of a microchannel section appropriately 3D printed to better visualize the trapping system. (**d**) Schematic of the microchannel used to trap a single *C. elegans* worm. Close up views of the trapping pillars and the central part of the microfluidic structure are shown in the insets. (**e**) Top view optical microscope picture of the pillars used to trap the head or tail of the *C. elegans*. (**f**) Top view optical microscope picture of the three microfluidic channels. The central microchannel traps the worm while the two lateral ones provide food to the trapped *C. elegans*. (**g**) SEM picture of a 3D printed channel section showing the trapping pillars. (**h**) SEM picture of a 3D printed channel section showing the central trapping microchannel and two lateral feeding microchannels.

**Design of the microchannels**

The microfluidic device has to gently handle the biological entity, and to accurately position and hold the sample in proximity of the detector microcoil, while providing a stable sealing over the duration of the experiments. In order to allow an easy alignment of the sample with the microcoil, the structural material of the microchannel has to be optically transparent. The geometry of the channel is designed to minimize static magnetic field inhomogeneities and maximize the detected NMR signal amplitude. We realized two microfluidic devices, designed for driving and trapping two different types of microscopic living entities in the most sensitive volume of the NMR integrated detector. These two designs are conceived, respectively, for NMR investigations of single tardigrade *Richtersius coronifer* (*Rc*) ova and of single *C. elegans* worms.

The design of the microchannel for the trapping of a single *Rc* ovum consists of a cylindrical inlet and outlet with a diameter of 240 μm. The microchannel gradually narrows to form a rectangular cross section with a height of 130 μm and a width of 140 μm at its centre. In correspondence to this constriction, two 10-μm-high pillars are designed to trap the ovum over the duration of the experiment (see Fig. 2a). A second, larger and taller pillar is implemented to prevent the loss of the ovum in case of accidental overpressure during the loading of the sample. The trapping of the ovum in the location of interest does not prohibit the flow in the channel, which is essential if we want to apply reverse flow to recover the sample or direct flow to refresh or change the medium. Fig. 2b shows an optical microscope picture of the top view of the central part of the microchannel. To better visualize the cross section of the microchannel and the system of pillars used to trap the ovum, a section of the microfluidic structure is 3D printed and inspected by scanning electron microscopy (SEM) (Fig. 2c).

The microfluidic chip designed for *C. elegans* is configured with one inlet and one outlet, featuring a main entrance that splits into three channels that re-join at the outlet (Fig. 2d). The central microchannel is designed to match tightly the size of an adult worm, being 60 μm wide, 50 μm high and 1.3 mm long. 50-μm-high pillars at the end of this channel are placed in a V-shaped arrangement at a distance of 12 μm, and used to block either the tail or the head of the *C. elegans* worm. The other two 55-μm-wide lateral channels are designed to deliver nutrients to the living trapped worm, being in fluidic connection with the two extremities of the central channel. To prevent the undesired trapping of other worms in the lateral feeding channels, one pillar is placed at their entrance (Fig. 2d). Top view optical microscope pictures of the trapping pillars and the central part of the microfluidic structure are shown respectively in Figs. 2e and 2f. Sections of the microfluidic device are 3D printed and inspected by SEM to better visualize the V-shaped arrangement of the pillars (Fig. 2g) and the part of the microfluidic structure formed of a central trapping microchannel and two lateral feeding microchannels (Fig. 2h).

In both designs the gradual narrowing of the channels facilitates the hydrodynamic trapping of the samples. The size and shape of the inlet and outlet parts match the capillaries used for the fluidic connection with the external pumps, limiting the presence of dead volumes (see details in ESI, section S.I). Similarly, the thickness of the separation layer gradually changes over the microchannel length, with a thickness of only 10 μm in proximity of the most sensitive region of the detector (i.e., above the excitation/detection microcoil). This strategy simultaneously implements the requirements of proximity for the maximization of the NMR signal amplitude and robustness of the structure.

**Fabrication of the microchannels**

The microchannels are fabricated via a two-photon polymerization technique using a high resolution 3D printer (Photonic Professional GT, Nanoscribe GmbH, Germany). This additive manufacturing technique provides a resolution better than 1 μm$^3$ in a more versatile manner compared to traditional microfabrication methods. The structures are patterned into IP-S photoresist, a negative tone cross-linking polymer proprietary to Nanoscribe GmbH, which is optically transparent and gives an extraordinary geometrical freedom in the design of the microchannels. An indium tin oxide (ITO) coated glass slide serves as a substrate. The refractive index difference between the two materials guarantees a facilitated detection of the interface. For the release of the microchannels, a 500-nm-thick dextran layer (Sigma-Aldrich, 31390-25G) is spin coated onto the substrate. This sacrificial layer is subsequently dissolved in water, enabling the release of the structures at the end of the fabrication process. A 25x objective (LCI Plan-Neofluar 25/0.8 Imm Korr DIC M27, Zeiss, Germany) is used in direct immersion mode into the photoresist. The refractive indices of the objective and the photoresist are matched to enable high spatial resolution. The voxel diameter is determined by a preceding test exposure, and the results are fed into the modelling software which imports Solidworks STL files to generate machine specific data. Using the 25x objective and 50 m/s writing speed, the voxel size is about (0.3x0.3x2) μm$^3$ (with a 63x objective the voxel size is about (0.15x0.15x0.45) μm$^3$). A droplet of IP-S resist is placed onto the substrate, in which the objective is immersed. The objective is fixed in space, while the positioning of the substrate is given by the combination of galvanometric MEMS mirrors and a piezoelectric unit. In the *xy* plane, the galvanometric mirrors travel within a 200 μm radius at each fixed piezo position. A slicing distance of 1 μm is defined to split the structure into equal distance horizontal planes which are set by the piezoelectric stage. This parameter is set to ensure proper overlapping and adhesion in between the horizontal planes. To decrease writing time, a scaffolding technique is implemented within the bulk volumes. A triangular support structure is used, with a 20 μm spacing between planes and a scaffolding wall thickness of 3 μm. The writing time for a single microfluidic structure is approximately 5 hours. To provide control over the resist-air interface, and to maintain mechanical resistance over an extended period of development, an 18 μm thick outer shell is defined, which is patterned as a bulk area. The accessible area of the galvanometric mirror is limited by the beam deflection and it is confined in a 200 μm radius circle. The block is a volume which can be written at a single *xy* piezoelectric stage position, only by moving the galvanometric mirrors and the piezoelectric stage in the *z* direction. The block shape has hexagonal shape to optimize the volume accessible from the galvanometric mirror and to facilitate stitching by having large neighbouring block surfaces. The blocks are written in a consecutive manner: a piezoelectric stage *xy* position is chosen, which gets exposed by the combination of the galvanometric mirrors and the piezoelectric *z* stage. When the block writing is finished, the piezoelectric stage moves to the next, neighbouring position and restarts exposure as before. A block shear angle of 13° is used in the *z* direction to avoid a shadowing effect, which occurs at overlapping block edges due to increased exposure. A block overlap of 2.5 μm enhances the stitching by reinforcing the adhesion between blocks. The block size is chosen to be *x*: 220 μm, *y*: 190 μm, *z*: 250 μm. Applying the shear angle and overlap, the total block dimension becomes *x*: 259.5 μm, *y*: 259 μm, *z*: 250 μm.

The laser power is 65 mW at the point of entering the objective. For writing the shell of the structure, the laser power is reduced to 42%. For the internal scaffolding, the laser power is increased to 50%, which increases the robustness at the cost of the spatial resolution. Following exposure, the objective is removed from the photoresist, and the substrate is placed into PGMEA (Sigma-Aldrich, 484431) for development. The substrate is positioned in a way that the channels stand vertically. A continuous stirring in the beaker enhances the removal of developed photoresist residues from inside the channel. The development duration is typically 6 hours. When development is finished, the substrate is immersed into isopropanol in an identical configuration for one hour. A second rinse step, with ultra-high purity isopropanol (99.99 %+) is used to further clean the inside of the channels, with a duration of 30 minutes. Finally, the substrate is left to naturally dry in vertical position.

**Assembly of the 3D printed microchannels with the single chip integrated CMOS detector**

In order to tightly hold the high resolution 3D printed microfluidic chips in contact with the single-chip integrated NMR detector and to connect its microchannels to external computer controlled syringe pumps, we fabricated an interfacing microfluidic structure (see details in sections S.I and S.II of the ESI). The positioning of the 3D printed microfluidic chip over the integrated microcoil is performed under a standard optical microscope (MZ8, Leica, Germany).

**Single chip integrated detector and NMR experiments**

The single chip integrated NMR detector, realized with a 130 nm CMOS technology from STMicroelectronics, consists of a RF power amplifier, a RF low-noise preamplifier, a frequency mixer, an audio-frequency (AF) amplifier, transmit/receive switches, and an excitation/detection microcoil with an external diameter of 150 µm. A detailed description of the single chip integrated NMR detector is reported in Ref. [33]. All NMR experiments performed in this work consist of a RF pulsed excitation (about 300 MHz, pulse duration $\tau$ of a few µs) immediately followed by a detection time $T_D$ (50 ms to 10 s). This simple excitation/detection sequence is repeated $n$ times (1 to $10^6$) with a repetition time $T_R$ (50 ms to 10 s). A schematic of the NMR experimental set-up is shown in section S.III of the ESI.

***C. elegans* preparation and microfluidic manipulation**

The *C. elegans* wild type worms are cultured at 20 °C on nematode growth media (NGM) 90 mm Petri dishes seeded with *Escherichia coli* strain OP50. Worms are provided by the Caenorhabditis Genetics Center (University of Minnesota). HT115 *E. coli* bacteria are grown in Luria Broth (LB) with 100 µg/mL ampicillin and 12.5 µg/mL tetracycline overnight in a thermal shaker at 37 °C. The following day, 50 µL of the confluent bacterial cultures are used to inoculate freshly prepared LB medium containing only ampicillin. The new cultures are grown until reaching an optical density between 0.6 and 0.8, and 90 µL are used for seeding the experimental plates.
The microfluidic chip and tubes are first filled with S medium (prepared following the protocol reported in Ref. 70) using the fluidic setup described in section S.II of the ESI. A worm of desired size is transferred, using a worm picker, from the agarose plate to an S medium reservoir, from which it is sucked up in a tube connected to the device. Afterwards, with a flow of 500 nL/s, the worm is injected in the microfluidic chip from the inlet. In order to insert the animal in the trapping channel, a flow of 1 µL/s is applied. Once trapped, a gentle flow of 50 nL/s of *E. coli* bacteria is used to replace the S medium in the chip and therefore provide nutrients to the worm through the lateral channels. After observing the pharyngeal pumping in the worm, which confirms that it is eating properly, the flow is stopped and the tubes are clamped. The whole operation, and the following NMR experiments, are performed at 20 °C.

***Rc* ova preparation and microfluidic manipulation**

*Rc* ova were extracted from a moss sample collected in Öland (Sweden) by washing the substrate, previously submerged in water for 30 min, on sieves under tap water and then individually picking up eggs with a glass pipette under a dissecting microscope. The ova were shipped within 24 hours in sealed Eppendorf tubes with water and subsequently stored at -20 °C before use. The ova are first transferred from the Eppendorf into a Petri dish prepared with 3% $H_2O$-based agarose gel.
The microfluidic chip and tubes are first filled with the chosen medium ($D_2O$ or $H_2O$ according to the experiment). A single *Rc* ovum is transferred from the agarose plate to the medium reservoir using a manipulation pipette (Vitrolife, Sweden). The single ovum is then released into the tube connected to the microfluidic system directly from the pipette. Later, with a flow of 500 nL/s, the ovum is injected in the microfluidic chip through the inlet. In order to place the ovum on top of the two trapping pillars, a flow of 3 µL/s is applied. Once trapped, the flow is stopped and the tubes are clamped. The whole operation, and the following NMR experiments, are performed at 20 °C.

## Results

### Spectroscopy of liquid samples

In Fig. 3, we show a collection of $^1$H spectra obtained from liquid samples (water and lactic acid in water) at 7.05 T (300 MHz) contained in the *C. elegans*-dedicated microchannels shown in Fig. 2d. These measurements are performed to characterize the spectral resolution limits of the setup and to measure the spin sensitivity of the detector. All chemical shifts are expressed in ppm deviation from the resonance frequency of tetramethylsilane (TMS). Since this standard reference compound is not present in our samples, we assigned a chemical shift of 4.8 ppm to the peak of water (present in all the investigated samples). The excitation pulse length used in the reported experiments ($\tau$ = 3.5 µs) corresponds to the experimental condition of maximum sensitivity, in good agreement with the value of 3.7 µs computed with simulations via sensitivity maps (Fig. 1b).

Fig. 3a shows the $^1$H NMR spectrum of pure water (Sigma-Aldrich, 270733) after averaging 1000 scans. The lorentzian fit of the data indicates a linewidth of 2 Hz FWHM. The baseline width, defined as the peak width at 0.55% height of the peak of water, is 24 Hz. These spectral resolutions are systematically achieved in six separated experiments employing six different microchannels with the same nominal design. Fig. 3b shows the $^1$H NMR spectrum of 1.3 M of lactic acid (Sigma-Aldrich, L1750) in pure water after averaging over 10800 scans. The FWHM in this spectrum is also of 2 Hz. As shown in the inset, the achieved spectral resolution is sufficient to show the 7 Hz J-splitting within each of the two chemically shifted signals of the lactic acid. The two peaks at about 1.3 ppm arise from the $^1$H nuclei in the $CH_3$ group, J-split by the $^1$H nucleus of the nearby CH group. The four peaks at about 4.08 ppm arise from the $^1$H nucleus of the CH group, and show the 1:3:3:1 J-split due to the $^1$H spins in the $CH_3$ group.[71] Better spectral resolutions (down to 0.6 Hz) have been reported in literature,[9, 28, 37, 43] but with probes having a worse spin sensitivity. In Fig. S4 of the ESI, we report the spectrum of pure water obtained in the *Rc* ova-dedicated microchannel. Contrarily to the case of *C. elegans*-dedicated microchannels (Fig. 2d), in this microchannel we inserted pillars in close proximity to the sensed volume, which are necessary to trap the *Rc* ova (Fig. 2a). The mismatches in susceptibility are larger than in the microchannel designed for the trapping of the *C. elegans*, resulting in an experimentally measured spectral resolution of 10 Hz (i.e., 5 times worse with respect to the other design). The approach proposed in this work (i.e., high spatial resolution 3D printed microchannels combined with single-chip CMOS integrated detectors) allows for sub-nL NMR spectroscopy with spectral resolutions consistently limited by the specific design of the microfluidic structure, with an experimentally demonstrated spectral resolution of 0.007 ppm for one of the two implemented designs. We used the NMR spectra obtained with the water sample also to calibrate the sensitivity of the detector in combination with the microfluidic channel. A value of 2.5x10$^{13}$ spins/Hz$^{1/2}$ is experimentally obtained for both microchannels. This value of sensitivity differs by a factor of 1.7 with respect to the experimental value found when the samples are placed in direct contact with the microcoil.[33, 41] Overall, this is in good agreement with what is computed via the sensitivity maps shown in Fig. 1b, where we estimate that a 10 µm separation layer reduces the sensitivity of the experiment by a factor of 1.4. The measured time-domain spin sensitivity corresponds to a limit of detection (LOD) at 300 MHz of approximately 600 pmol s$^{1/2}$ of $^1$H nuclei in pure water ($T_1 \cong T_2 \cong 3$ s) with an effective transversal relaxation time $T_2^* \cong 0.15$ s (i.e., a 2 Hz FWHM line) and repetition time $T_R$ = 4 s.

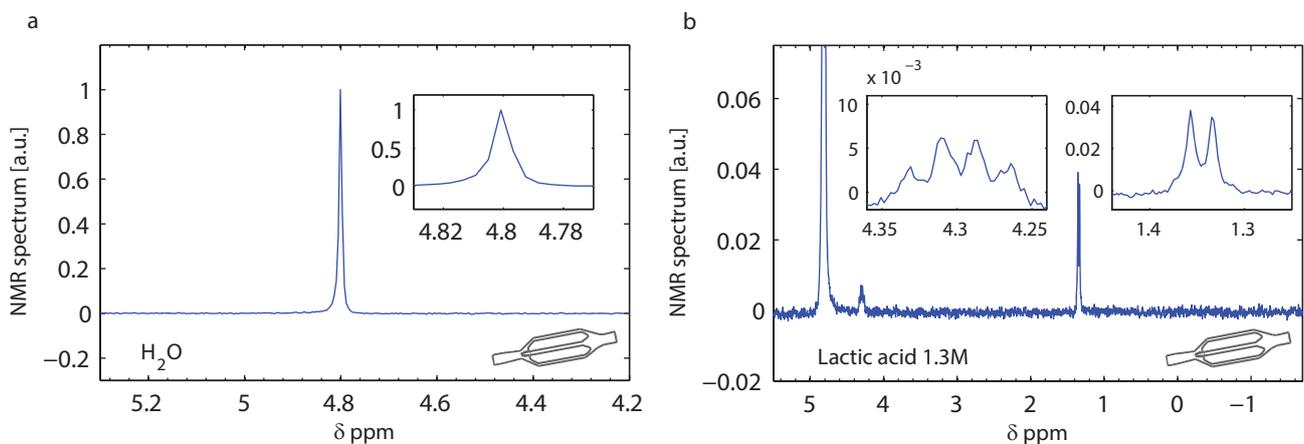

**Fig. 3**: $^1$H NMR measurements on pure water and lactic acid in water performed at 7.05 T (300 MHz). The spectra are the real parts of the Fast Fourier Transform (FFT) of the time-domain NMR signals. Notations: *V* is the active volume, *Avg* number of averaged measurements, $T_R$ is the repetition time, $\tau$ is the pulse length, $T_m$ is the matching filter decay time constant. (**a**) $^1$H spectrum of water: $V \cong 100$ pL, *Avg* = 1000, $T_R$ = 4 s, $\tau$ = 3.5 µs, $T_m = \infty$. (**b**) $^1$H spectrum of 1.3 M lactic acid in $H_2O$: $V \cong 100$ pL, *Avg* = 10800, $T_R$ = 2 s, $\tau$ = 3.5 µs, $T_m$ = 500 ms.

**Spectroscopy of single intact biological samples**

Figure 4 shows $^1$H NMR spectra obtained at 7.05 T (300 MHz) from intact biological samples. Figures 4a and 4b show NMR spectra obtained from a single *Rc* ovum. Figures 4c and 4d show the NMR spectrum of a subsection of a single *C. elegans* worm. *Rc* ova have typical volumes of 500 pL, whereas adult *C. elegans* worms have typical volumes of 5 nL. The most sensitive region of the NMR integrated detector corresponds to a deformed ellipsoid of about 300 pL (Fig. 1b). In the experiments, the sensed portion of the microorganisms is given by the intersection of the sensitive region with the volume of the sample. Defining the active volume as the fraction of the sample that contributes to the 70% of the total signal and considering the geometries in play and their position with respect to the microcoil, we estimate active volumes of about 250 pL in the case of the *Rc* ovum and 100 pL in the case of the *C. elegans* worm. The chemical shifts scale is obtained assigning a chemical shift of 4.8 ppm to the peak of water contained in the samples under investigation.

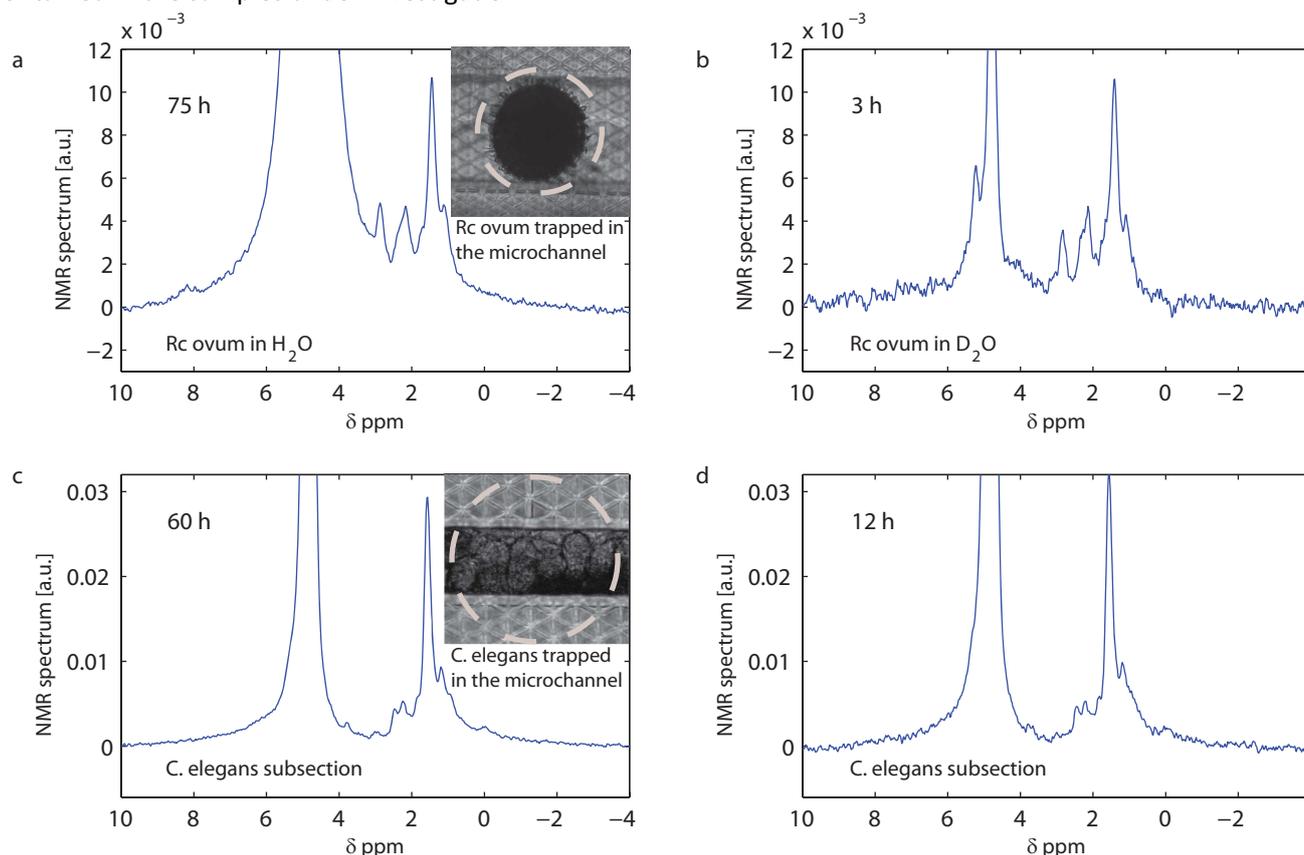

**Fig. 4**: $^1$H NMR measurements at 7.05 T (300 MHz). See definition of notations in Fig. 3. The dashed circles indicate the 150 μm outer diameter of the integrated microcoil. (**a**) $^1$H spectrum of a single *Rc* tardigrade ovum in $H_2O$: $V \sim 150$ pL, Avg = 710400, $T_R$ = 200 ms, $\tau$ = 3.5 μs, $T_m$ = 30 ms. (**b**) $^1$H spectrum of a single *Rc* tardigrade ovum in $D_2O$: $V \cong 150$ pL, Avg = 28416, $T_R$ = 200 ms, $\tau$ = 3.5 μs, $T_m$ = 30 ms. (**c**) $^1$H spectrum of a single *C. elegans* subsection: $V \cong 100$ pL, Avg = 568320, $T_R$ = 200 ms, $\tau$ = 3.5 μs, $T_m$ = 30 ms. (**d**) $^1$H spectrum of a single *C. elegans* subsection: $V \cong 100$ pL, Avg = 113664, $T_R$ = 200 ms, $\tau$ = 3.5 μs, $T_m$ = 30 ms.

Fig. 4a shows the NMR spectrum obtained from a single *Rc* ovum in $H_2O$ after averaging over 75 hours. With a FWHM linewidth of about 0.1 ppm (35 Hz) and a baseline width of about 3 ppm (1 kHz), the water peak can overlap significantly with eventual small nearby signals. To suppress the water peak which originate from the carrying medium, we measured the Rc ova also in $D_2O$. The same *Rc* ovum is used to perform the measurements in $H_2O$ and $D_2O$ shown in Figs. 4a and 4b. In $D_2O$ at least one additional peak is more clearly visible. Thanks to the fluidic interface (see section S.II of the ESI), the liquid is easily exchanged in-between the two experiments without repeating the trapping procedure. Due to the relatively low SNR (caused by the very small number of spins contained in the sample) and the relatively poor spectral resolution, a detailed proton peak assignment is currently impossible. Nevertheless, some information can be obtained from previous studies on *Xenopus laevis*[56, 72] and hen eggs[73]. The spectra shown in Figs. 4a and 4b indicate that we can associate the nature of the dominant peaks in the *Rc* ovum (i.e. the peaks at 0.9, 1.3, 2.1, 2.8 ppm) to the yolk lipid content of the sample.

Fig. 4c shows the NMR spectrum obtained from a single *C. elegans* subsection, after averaging over 60 hours. After trapping the worm using S medium, a gentle flow of S medium and *E.coli* (1.5 x 10$^9$ cell/mL) is applied to provide food to the *C. elegans*. At this point the flow is stopped and the fluidic connection clamped for sealing. The

*C. elegans* goes through a squeezing process when entering the microchannel, whose size tightly matches the worm. It follows that, in good approximation, only the worm is present in the active area of the detector. For this experiment, we decided to align the subsection of the worm containing ova with the microcoil. As shown in Fig. 4c (inset), most of the sensitive region is indeed occupied by about eight eggs contained in *C. elegans* abdomen. The resulting spectrum has features which are similar to those measured in the *Rc* eggs. The peaks assignment is challenging due to the reduced size of the samples combined with the relatively poor spectral resolution. The peaks observed at about 0.9, 1.3, 2.1, 2.8 ppm may be mainly associated to yolk lipids. Some qualitative information can be obtained from previous $^1$H HRMAS NMR studies on a collection of *C. elegans*[49] and on a single worm[51]. A prominent signal arises at a chemical shift of about 3.9 ppm. As shown in the Fig. S6 of the ESI, the same resonance appears in the spectrum of *E.coli* in S medium, which is used as a base to feed the worm during the experiment. Therefore, we suspect that such resonance may result from the ingestion of nutrients operated by the microorganism.

In both *Rc* and *C. elegans* spectra, the linewidths are broader than the ones measured with liquid samples (as shown in Fig.3 for pure water and lactic acid). Previous studies suggest that the spectral resolution may be limited by microscopic constituents in the samples, which introduce susceptibility mismatches which are difficult to compensate by field shimming. As shown above, the linewidths observed in this study are of 35 Hz FWHM for the *Rc* ovum and 30 Hz FWHM for the *C. elegans*. For these biological entities, measured with a repetition time $T_R = 200$ ms and having linewidths of 30 Hz FWHM (i.e., $T_2^* \cong 0.01$ s), the LOD is about 600 pmol s$^{1/2}$ of $^1$H nuclei, which corresponds to 6 pmol of $^1$H nuclei in an averaging time of 3 hours (see further experimental details in section S.V of the ESI).

Thanks to the freedom in the design of the microchannels, the stable trapping, and the robustness of the fluidic interface, the duration of our experiments depends only on the viability of the biological sample under investigation. In fact, we achieved measuring times as long as two weeks on both single microorganisms. The successful implementation of long measurements allows to improve the signal-to-noise ratio and to perform continuous observations on the same single entity on a time scale that can be comparable to the biological development of the sample.

## Conclusions and outlook

NMR experiments on sub-nL volumes are hindered by the limited sensitivity of the detector and the difficulties in positioning and holding such small samples in proximity of the excitation/detection microcoils. In this work, we show the first examples of NMR experiments on liquids and biological entities immersed in liquids having active volumes down to 100 pL. These measurements are enabled by the successful design and fabrication of high spatial resolution 3D printed microfluidic structures, specifically conceived to guide and confine sub-nL samples in the sub-nL most sensitive volume of a single-chip integrated NMR detector. The obtained results indicate a promising route for NMR studies at the single unit level of important biological entities having sub-nL volumes, which include, e.g., the eggs of several mammalians.[59]

The microfluidic structures are fabricated using a two-photon polymerization 3D printing technique having a resolution better than better than 1 µm$^3$. Compared to conventional clean-room microfabrication techniques, high resolution 3D printing techniques are currently not appropriate for an efficient large scale production, but represent a very promising solution for the rapid prototyping of simple and complex microfluidic systems[74-76] for the handling, feeding and trapping of biological samples having different geometry, size, and fluidic behaviour. More specifically, in this work we show that high resolution 3D printing is well suited for the microfabrication of microfluidic structures dedicated to NMR spectroscopy studies of sub-nL samples. Thanks to the fabrication of a robust separation layer of only 10 µm between the sample and the excitation/detection microcoil, the sample can be trapped in a liquid environment for several days very close to most sensitive volume of the detector. The obtained effective spin sensitivity of 2.5·10$^{13}$ spin/Hz$^{1/2}$ is sufficient for the detection of highly concentrated endogenous compounds in the sub-nL volumes of the investigated biological entities (i.e., single *Rc* ova and portion of *C. elegans* worms) in experimental times of a few hours. In the measurements on the intact biological samples we achieved spectral resolutions of approximately 0.1 ppm, due to mismatches in the susceptibility caused by the sample itself, which we have not managed to shim. In the microchannels designed to minimize magnetic field non-uniformities we obtained spectral resolutions down to 0.007 ppm (i.e., about 2 Hz at 7 T) in 100 pL liquid samples.

The combination of CMOS integrated NMR detectors with high spatial resolution 3D printed microfluidic structures is compatible with the implementation of arrays of miniaturized probes, which would enable simultaneous studies on a large number of single biological entities in the same magnet. This would, in turn, allow for systematic investigation of the heterogeneity among individuals or among different subsections of a single organism (such as a *C. elegans* worm), as well as of their response to different environmental conditions and drugs.

## Conflict of interest

There are no conflicts to declare.

## Acknowledgements

We thank Vincenzo Sorrentino for kindly providing the *C. elegans* worm plates.

## References

1. A. Abragam, *The Principles opf Nuclear Magnetic Resonance*, Clarendon Press, Oxford, 1961.
2. P. T. Callaghan, *Principles of nuclear magnetic resonance microscopy*, Clarendon Press Oxford, 1991.
3. R. R. Ernst, G. Bodenhausen and A. Wokaun, *Principles of nuclear magnetic resonance in one and two dimensions*, 1991.
4. I. L. Pykett, *Sci Am*, 1982, **246**, 78-88.
5. D. W. McRobbie, E. A. Moore, M. J. Graves and M. R. Prince, *MRI from Picture to Proton*, Cambridge university press, 2017.
6. I. Tkac, Z. Starcuk, I. Y. Choi and R. Gruetter, *Magnet Reson Med*, 1999, **41**, 649-656.
7. I. Tkac, P. Andersen, G. Adriany, H. Merkle, K. Ugurbil and R. Gruetter, *Magnet Reson Med*, 2001, **46**, 451-456.
8. J. Pfeuffer, I. Tkac, S. W. Provencher and R. Gruetter, *Journal of Magnetic Resonance*, 1999, **141**, 104-120.
9. D. L. Olson, T. L. Peck, A. G. Webb, R. L. Magin and J. V. Sweedler, *Science*, 1995, **270**, 1967-1970.
10. T. L. Peck, R. L. Magin and P. C. Lauterbur, *J Magn Reson Ser B*, 1995, **108**, 114-124.
11. A. G. Webb, *Progress in Nuclear Magnetic Resonance Spectroscopy*, 1997, **31**, 1-42.
12. M. E. Lacey, R. Subramanian, D. L. Olson, A. G. Webb and J. V. Sweedler, *Chem. Rev.*, 1999, **99**, 3133-+.
13. D. Seeber, R. Cooper, L. Ciobanu and C. Pennington, *Rev. Sci. Instrum.*, 2001, **72**, 2171-2179.
14. G. Boero, J. Frounchi, B. Furrer, P.-A. Besse and R. Popovic, *Rev. Sci. Instrum.*, 2001, **72**, 2764-2768.
15. K. R. Minard and R. A. Wind, *Journal of Magnetic Resonance*, 2002, **154**, 336-343.
16. C. Massin, F. Vincent, A. Homsy, K. Ehrmann, G. Boero, P. A. Besse, A. Daridon, E. Verpoorte, N. F. de Rooij and R. S. Popovic, *Journal of Magnetic Resonance*, 2003, **164**, 242-255.
17. L. Ciobanu and C. Pennington, *Solid state nuclear magnetic resonance*, 2004, **25**, 138-141.
18. K. Yamauchi, J. W. G. Janssen and A. P. M. Kentgens, *Journal of Magnetic Resonance*, 2004, **167**, 87-96.
19. D. Sakellariou, G. Le Goff and J.-F. Jacquinot, *Nature*, 2007, **447**, 694-697.
20. Y. Maguire, I. L. Chuang, S. G. Zhang and N. Gershenfeld, *P Natl Acad Sci USA*, 2007, **104**, 9198-9203.
21. P. J. M. van Bentum, J. W. G. Janssen, A. P. M. Kentgens, J. Bart and J. G. E. Gardeniers, *Journal of Magnetic Resonance*, 2007, **189**, 104-113.
22. K. Ehrmann, N. Saillen, F. Vincent, M. Stettler, M. Jordan, F. M. Wurm, P. A. Besse and R. Popovic, *Lab Chip*, 2007, **7**, 373-380.
23. H. G. Krojanski, J. Lambert, Y. Gerikalan, D. Suter and R. Hergenroder, *Anal Chem*, 2008, **80**, 8668-8672.
24. M. Weiger, D. Schmidig, S. Denoth, C. Massin, F. Vincent, M. Schenkel and M. Fey, *Concepts in Magnetic Resonance Part B: Magnetic Resonance Engineering*, 2008, **33**, 84-93.
25. J. Anders, G. Chiaramonte, P. SanGiorgio and G. Boero, *Journal of Magnetic Resonance*, 2009, **201**, 239-249.
26. S. Leidich, M. Braun, T. Gessner and T. Riemer, *Concepts in Magnetic Resonance Part B: Magnetic Resonance Engineering*, 2009, **35**, 11-22.
27. M. H. Lam, M. A. Homenuke, C. A. Michal and C. L. Hansen, *J Micromech Microeng*, 2009, **19**, 095001.
28. J. Bart, J. W. G. Janssen, P. J. M. van Bentum, A. P. M. Kentgens and J. G. E. Gardeniers, *Journal of Magnetic Resonance*, 2009, **201**, 175-185.
29. K. Kratt, V. Badilita, T. Burger, J. G. Korvink and U. Wallrabe, *J Micromech Microeng*, 2010, **20**.
30. J. Anders, P. SanGiorgio and G. Boero, *Journal of Magnetic Resonance*, 2011, **209**, 1-7.
31. V. Badilita, R. C. Meier, N. Spengler, U. Wallrabe, M. Utz and J. G. Korvink, *Soft Matter*, 2012, **8**, 10583-10597.
32. R. C. Meier, J. Hofflin, V. Badilita, U. Wallrabe and J. G. Korvink, *J Micromech Microeng*, 2014, **24**.
33. M. Grisi, G. Gualco and G. Boero, *Rev. Sci. Instrum.*, 2015, **86**.
34. A. Kalfe, A. Telfah, J. Lambert and R. Hergenroder, *Anal Chem*, 2015, **87**, 7402-7410.
35. G. Finch, A. Yilmaz and M. Utz, *Journal of Magnetic Resonance*, 2016, **262**, 73-80.
36. J. Anders, J. Handwerker, M. Ortmanns and G. Boero, *Journal of Magnetic Resonance*, 2016, **266**, 41-50.
37. N. Spengler, J. Höfflin, A. Moazenzadeh, D. Mager, N. MacKinnon, V. Badilita, U. Wallrabe and J. G. Korvink, *PloS one*, 2016, **11**, e0146384.
38. K. C. Tijssen, J. Bart, R. M. Tiggelaar, J. H. Janssen, A. P. Kentgens and P. J. M. van Bentum, *Journal of Magnetic Resonance*, 2016, **263**, 136-146.

39. N. T. Duong, Y. Endo, T. Nemoto, H. Kato, A. K. Bouzier-Sore, Y. Nishiyama and A. Wong, *Anal Methods-Uk*, 2016, **8**, 6815-6820.
40. R. Kamberger, A. Moazenzadeh, J. G. Korvink and O. G. Gruschke, *J Micromech Microeng*, 2016, **26**.
41. M. Grisi, F. Vincent, B. Volpe, R. Guidetti, N. Harris, A. Beck and G. Boero, *Scientific Reports*, 2017, **7**.
42. J. Bart, A. J. Kolkman, A. J. Oosthoek-de Vries, K. Koch, P. J. Nieuwland, H. Janssen, P. J. M. van Bentum, K. A. M. Ampt, F. P. J. T. Rutjes, S. S. Wijmenga, H. Gardeniers and A. P. M. Kentgens, *J. Am. Chem. Soc.*, 2009, **131**, 5014-+.
43. J. A. Rogers, R. J. Jackman, G. M. Whitesides, D. L. Olson and J. V. Sweedler, *Applied Physics Letters*, 1997, **70**, 2464-2466.
44. H. Mamin, T. Oosterkamp, M. Poggio, C. Degen, C. Rettner and D. Rugar, *Nano letters*, 2009, **9**, 3020-3024.
45. C. Degen, M. Poggio, H. Mamin, C. Rettner and D. Rugar, *Proceedings of the National Academy of Sciences*, 2009, **106**, 1313-1317.
46. H. Mamin, M. Poggio, C. Degen and D. Rugar, *Nat. Nanotechnol.*, 2007, **2**, 301-306.
47. S. J. DeVience, L. M. Pham, I. Lovchinsky, A. O. Sushkov, N. Bar-Gill, C. Belthangady, F. Casola, M. Corbett, H. Zhang and M. Lukin, *Nat. Nanotechnol.*, 2015, **10**, 129-134.
48. J. Boss, K. Chang, J. Armijo, K. Cujia, T. Rosskopf, J. Maze and C. Degen, *Physical review letters*, 2016, **116**, 197601.
49. B. J. Blaise, J. Giacomotto, B. Elena, M.-E. Dumas, P. Toulhoat, L. Ségalat and L. Emsley, *Proceedings of the National Academy of Sciences*, 2007, **104**, 19808-19812.
50. B. J. Blaise, J. Giacomotto, M. N. Triba, P. Toulhoat, M. Piotto, L. Emsley, L. Ségalat, M.-E. Dumas and B. Elena, *Journal of proteome research*, 2009, **8**, 2542-2550.
51. A. Wong, X. Li, L. Molin, F. Solari, B. n. d. Elena-Herrmann and D. Sakellariou, *Anal Chem*, 2014, **86**, 6064-6070.
52. J. B. Aguayo, S. J. Blackband, J. Schoeniger, M. A. Mattingly and M. Hintermann, *Nature*, 1986, **322**, 190-191.
53. J. S. Schoeniger, N. Aiken, E. Hsu and S. J. Blackband, *J Magn Reson Ser B*, 1994, **103**, 261-273.
54. S. C. Grant, N. R. Aiken, H. D. Plant, S. Gibbs, T. H. Mareci, A. G. Webb and S. J. Blackband, *Magnet Reson Med*, 2000, **44**, 19-22.
55. S. C. Grant, D. L. Buckley, S. Gibb, A. G. Webb and S. J. Blackband, *Magnet Reson Med*, 2001, **46**, 1107-1112.
56. S. C. Lee, J. H. Cho, D. Mietchen, Y. S. Kim, K. S. Hong, C. Lee, D. M. Kang, K. D. Park, B. S. Choi and C. Cheong, *Biophys J*, 2006, **90**, 1797-1803.
57. S. C. Lee, D. Mietchen, J. H. Cho, Y. S. Kim, C. Kim, K. S. Hong, C. Lee, D. Kang, W. Lee and C. Cheong, *Differentiation*, 2007, **75**, 84-92.
58. C. H. Lee, J. J. Flint, B. Hansen and S. J. Blackband, *Scientific Reports*, 2015, **5**, 11147.
59. R. Flindt, *Amazing numbers in biology*, Springer Science & Business Media, 2006.
60. B. Sorli, J. F. Chateaux, M. Pitaval, H. Chahboune, B. Favre, A. Briguet and P. Morin, *Measurement Science and Technology*, 2004, **15**, 877-880.
61. D. A. Seeber, R. L. Cooper, L. Ciobanu and C. H. Pennington, *Rev. Sci. Instrum.*, 2001, **72**, 2171-2179.
62. C. Massin, C. Boero, F. Vincent, J. Abenhaim, P. A. Besse and R. S. Popovic, *Sensor Actuat a-Phys*, 2002, **97-8**, 280-288.
63. V. Demas, A. Bernhardt, V. Malba, K. L. Adams, L. Evans, C. Harvey, R. S. Maxwell and J. L. Herberg, *Journal of Magnetic Resonance*, 2009, **200**, 56-63.
64. V. Malba, R. Maxwell, L. B. Evans, A. E. Bernhardt, M. Cosman and K. Yan, *Biomed Microdevices*, 2003, **5**, 21-27.
65. L. O. Sillerud, A. F. McDowell, N. L. Adolphi, R. E. Serda, D. P. Adams, M. J. Vasile and T. M. Alam, *Journal of Magnetic Resonance*, 2006, **181**, 181-190.
66. B. H. Cumpston, S. P. Ananthavel, S. Barlow, D. L. Dyer, J. E. Ehrlich, L. L. Erskine, A. A. Heikal, S. M. Kuebler, I.-Y. S. Lee and D. McCord-Maughon, *Nature*, 1999, **398**, 51-54.
67. X. Zhou, Y. Hou and J. Lin, *AIP Advances*, 2015, **5**, 030701.
68. S. H. Park, D. Y. Yang and K. S. Lee, *Laser & Photonics Reviews*, 2009, **3**, 1-11.
69. D. I. Hoult and R. Richards, *Journal of Magnetic Resonance (1969)*, 1976, **24**, 71-85.
70. T. Stiernagle, *C. elegans*, 1999, **2**, 51-67.
71. S. G. Lloyd, H. D. Zeng, P. P. Wang and J. C. Chatham, *Magnet Reson Med*, 2004, **51**, 1279-1282.
72. J. V. Sehy, J. J. H. Ackerman and J. J. Neil, *Magnet Reson Med*, 2001, **46**, 900-906.
73. R. Jayasundar, S. Ayyar and P. Raghunathan, *Magnetic resonance imaging*, 1997, **15**, 709-717.
74. S. Waheed, J. M. Cabot, N. P. Macdonald, T. Lewis, R. M. Guijt, B. Paull and M. C. Breadmore, *Lab Chip*, 2016, **16**, 1993-2013.
75. A. K. Au, W. Huynh, L. F. Horowitz and A. Folch, *Angewandte Chemie International Edition*, 2016, **55**, 3862-3881.
76. C. M. B. Ho, S. H. Ng, K. H. H. Li and Y.-J. Yoon, *Lab Chip*, 2015, **15**, 3627-3637.

# Electronic Supplementary Informations (ESI)

# 3D printed microchannels for sub-nL NMR spectroscopy

E. Montinaro[a], M. Grisi[a], M. C. Letizia[a], L. Pethö[b], M. A. M. Gijs[a], R. Guidetti[c], J. Michler[b], J. Brugger[a], and G. Boero[a,*]

## S.I: Fluidic interface

In order to precisely position and tightly hold the high resolution 3D printed microfluidic chips in contact with the single-chip integrated CMOS detector (and to connect its microchannels to the external pump) an interfacing structure is fabricated. The 3D printed microfluidic chips are first glued on a 200 µm x 2 cm PMMA rod for support using a cyanoacrylate adhesive (ECS500, 3M). After, PMMA capillaries (Paradigm Optics, USA) are fitted in the inlet and the outlet of the 3D printed microchannels. Finally, the realized assembly shown in Fig. S1a is introduced into a 3D printed plastic holder (Fig. S2b) fabricated by a conventional stereolithographic 3D printer (Form+1, Formlabs, USA). The holder is patterned out of a photosensitive resin (Clear FLGPCL02, Formlabs, USA) and is constituted of two complementary parts to allow for the insertion of the fluidic assembly. A system of tubes is connected to the fluidic assembly through the PMMA capillaries and a casting of epoxy resin (Araldite) is performed to seal the fluidic system and to give robustness to the structure. The assembled fluidic interface (Fig. S1c) is manually brought into contact and aligned to the integrated excitation/detection microcoil under an optical microscope. Wax is used to maintain the interface in position. A central squared aperture in the holder provides visibility for the alignment. The printed circuit board (PCB), containing the single-chip integrated NMR detector, includes two elongated holes to allow for sufficient free planar movement for the microchannel-to-microcoil alignment. With this system, we repeatedly achieved a good seal, easy handling of the liquids, and a positioning of the sample within 10 µm precision.

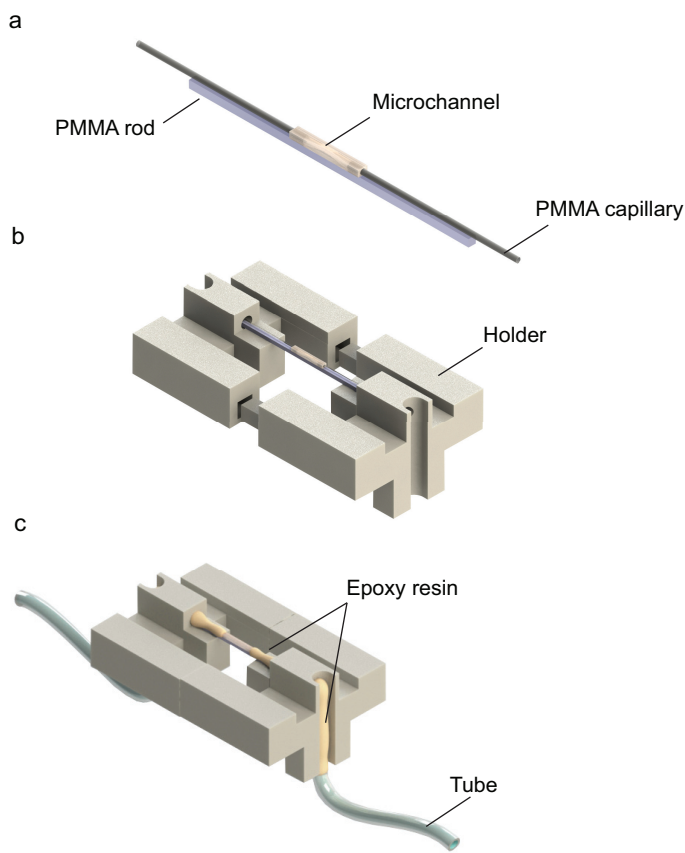

**Fig. S1. Illustration of the fluidic interface.** (**a**) The microfluidic channel is attached to a PMMA rod and connected to PMMA capillaries to create a fluidic assembly. (**b**) The fluidic assembly is mounted in the holder. (**c**) The Micro-to-macro interface is completed by connecting tubes to the PMMA capillaries and performing the casting of epoxy resin to create the sealing and give robustness to the fluidic system.

## S.II: Fluidic set-up for sample loading and trapping

Figure S2 shows an illustration of the fluidic set-up. The 3D printed microfluidic chip (together with its fluidic interface) is placed into an inverted microscope (Axio Observer, Zeiss, Germany) equipped with a High-Power LED Illumination system (precisExcite, Visitron, Germany) for brightfield imaging. The microscope is equipped with a motorized *xyz* stage that has a piezo controller for *z* displacement (ASI, Visitron, Germany). The microfluidic operations are controlled using syringe pumps and its software (Nemesys, Cetoni GmbH, Germany).

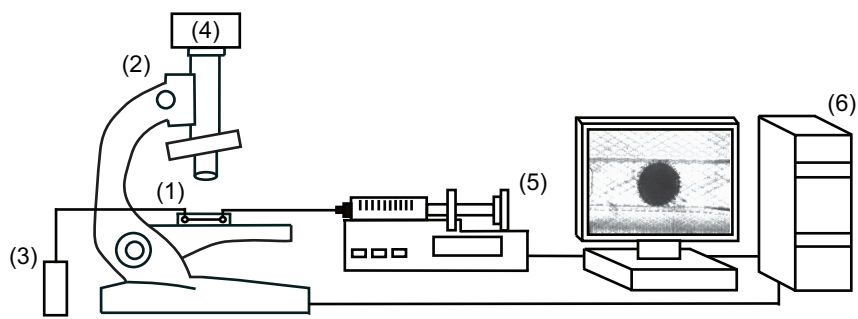

**Fig. S2. Schematic of the fluidic set-up used for the loading and trapping of biological samples.** (**1**) Microfluidic chip and its fluidic interface. (**2**) Inverted microscope (Axio Observer, Zeiss, Germany), High-Power LED Illumination system (precisExcite, Visitron, Germany) for brightfield imaging. (**3**) Liquid waste. (**4**) High resolution digital camera (ORCA-ER C4742-80, Hamamatsu, Japan). (**5**) Syringe pumps (Nemesys, Cetoni GmbH, Germany). (**6**) Desktop computer.

## S.III: NMR set-up

Figure S3 shows an illustration of the NMR experimental set-up. The single-chip CMOS integrated NMR detector is glued onto a printed circuit board (PCB) and electrically connected by wire bonding. A single RF generator is used to provide the RF signal for the transmission (Tx) and as local oscillator (LO) for the on-chip frequency down-conversion. The frequency down-converted and amplified signal and the output of the single-chip NMR detector is further amplified by an external amplifier and sent to a multifunctional board for acquisition. The PCB is inserted in the 54 mm room temperature bore of a 7.05 T superconducting magnet (Bruker, 300 MHz). The 3D printed microchannel lies along the direction of the static magnetic field to reduce the effect of susceptibility mismatches. In this configuration, the NMR linewidth is about a factor of two narrower with respect to the configuration where the channel is perpendicular to the static magnetic field.

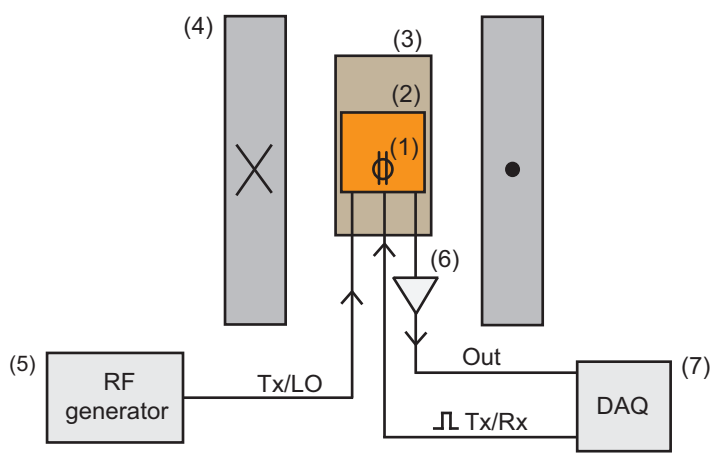

**Fig. S3. Schematic of the electronics setup used for the NMR measurements**. (**1**) Integrated excitation/detection coil interfaced with the 3D printed microchannel through a micro-to-macro fluidic interface. (**2**) Single chip NMR detector (see details in Ref. 1). (**3**) Printed circuit board (PCB). (**4**) Superconductive magnet (Bruker, 7 T). (**5**) RF source (MG3633A, Anritsu; Japan). (**6**) AF amplifier (SRS560, Stanford Research Systems, USA) (**7**) Multifunctional board (PCIe-6259, National Instruments, USA) for the generation of Tx/Rx switching pulse and signal acquisition.

## S.IV: ¹H NMR spectrum of H$_2$O in a microchannel designed for ovum trapping

Figure S4 shows the ¹H NMR spectrum of water after averaging over 2 hours. The NMR experiment is performed using the 3D printed microchannel designed for the handling and the trapping of a single *Rc* ovum, interfaced with the holding structure described in S.I. The chemical shift is expressed in ppm, assigning a chemical shift of 4.8 ppm to the peak of water. The excitation pulse length (3.5 µs) used in the reported experiments corresponds to the experimental condition of maximum signal-to-noise ratio. The spectrum is obtained at 7.05 T (300 MHz) and is normalized to the amplitude of the peak of water. The linewidth, defined as spectral peak width at half-maximum and determined through the fit of a Lorentzian curve to the data, is equal to 10 Hz FWHM. The baseline width, defined as the peak width at 0.55% height of the peak of water, is equal to 75 Hz.

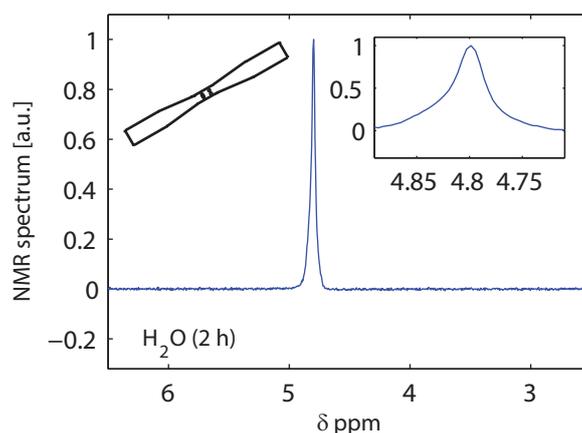

**Fig. S4. ¹H spectrum of H$_2$O.** NMR measurements performed at 7.05 T (300 MHz), using the 3D printed microchannel designed for the trapping of a single ovum. The spectra are the real parts of the Fast Fourier Transform (FFT) of the time-domain NMR signals. Notations: *V* is the active volume, *Avg* number of averaged measurements, $T_R$ is the repetition time, τ is the pulse length, $T_m$ is the matching filter decay time constant. $V \cong 250$ pL, *Avg* = 1800, $T_R$ = 2s, τ = 3.5 µs, $T_m = \infty$.

## S.V: Optimization of the repetition time

Figure S5 shows the $^1$H NMR spectra obtained from a single *Rc* ovum in H$_2$O, with a repetition time $T_R$ of 2 s (Fig. S5a), 200 ms (Fig. S5b) and 50 ms (Fig. S5c) after averaging over 12 hours, all obtained with the same pulse length τ = 3.5 μs. This pulse length maximize the amplitude of the signal at 1.3 ppm for $T_R$ = 2 s and longer. For all spectra the signal amplitude is normalized to the amplitude of the peak at 1.3 ppm obtained with a $T_R$ of 2 s. Reducing the repetition time $T_R$ from 2 s to 200 ms, the amplitudes of all observed peaks, water included, do not change significantly (in our previous study of the *Rc* ova,[2] we used a non-optimum repetition time of 2 s). Such faster repetition rate allow for a reduction of the experimental time of a factor 3.3 for the same SNR. A detailed study of the spin-lattice (and spin-spin) relaxation times of the $^1$H nuclei contained inside the different compounds in the investigated biological samples will be reported elsewhere. Due to the weak signal amplitudes and the highly inhomogeneous RF excitation field $B_1$, these investigations requires very long averaging time and computational care.

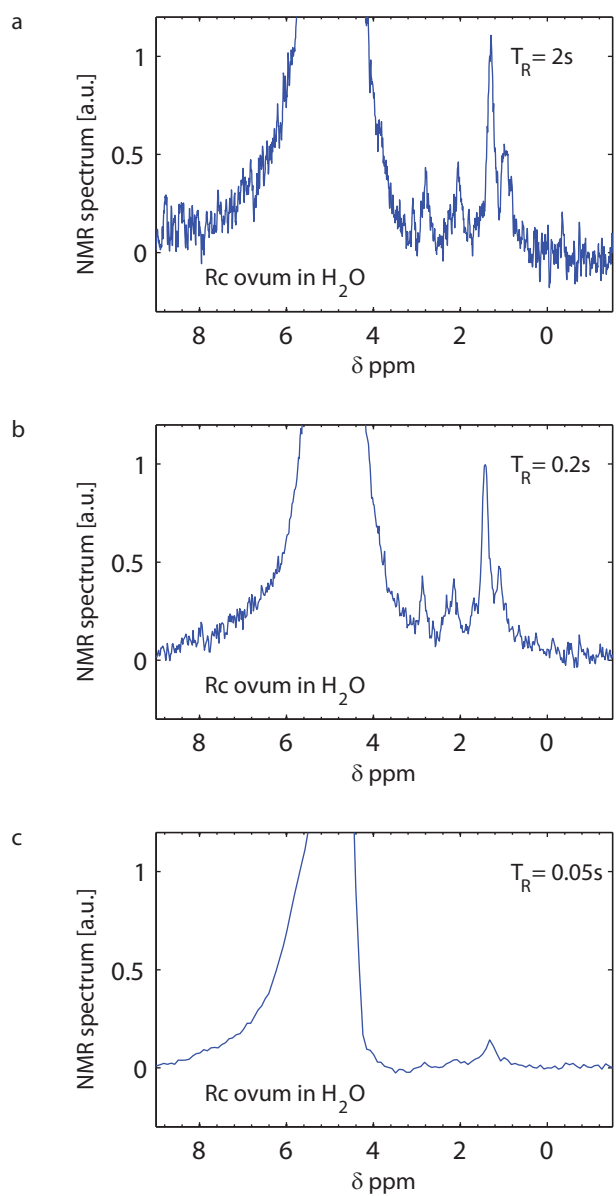

**Fig. S5. NMR measurements of a single *Rc* tardigrade ovum in H$_2$O at different repetition times.** NMR measurements performed at 7 T. See definition of notation in Fig. S4. **(a)**: $V \cong$ 210 pL; *Avg* = 21600; $T_R$ = 2 s, τ = 3.5 μs, $T_m$ = 30 ms. **(b)**: $V \cong$ 210 pL; *Avg* = 216000; $T_R$ = 200 ms, τ = 3.5 μs, $T_m$ = 30 ms. **(c)**: $V \cong$ 210 pL; *Avg* = 864000; $T_R$ = 50 ms, τ = 3.5 μs, $T_m$ = 30 ms.

## S.VI: ¹H NMR spectrum of *E.coli* in S medium

Figure S6 shows the ¹H NMR spectrum of *E. coli* in S medium (prepared following the protocol reported in Ref. 3 at a concentration of $1.5 \times 10^9$ cell/mL, obtained after 11 hours of averaging. This solution is used to feed the *C. elegans*. The signal acquired from the *E. coli* solution, ingested by the worm during the experiment, may contribute to the NMR spectrum obtained from the *C. elegans* subsection. All chemical shifts are expressed in ppm deviation from the resonance frequency of tetramethylsilane (TMS). Since this reference compound is not present in our samples, we assigned a chemical shift of 4.8 ppm to the peak of water, thus determining the chemical shifts of the other peaks. The excitation pulse length (3.5 μs) corresponds to the experimental condition of maximum signal-to-noise ratio. The spectrum is obtained using the 3D printed microchannel designed for the trapping of a single *C. elegans* and it is normalized to the amplitude of the peak of water.

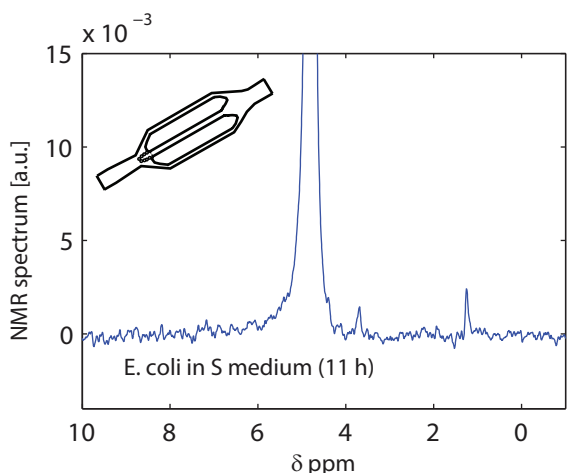

**Fig. S6. ¹H spectrum of *E.coli* in S medium.** NMR measurement performed at 7 T, using the 3D oriented microchannel designed for the trapping of a single *C. elegans*. See definition of notation in Fig. S4. $V \cong 100$ pL; $Avg$ = 19800; $T_R$ = 2 s, $\tau$ = 3.5 μs, $T_m$ = 60 ms.

## References

1. M. Grisi, G. Gualco and G. Boero, *Rev. Sci. Instrum.*, 2015, **86**, 044703.
2. M. Grisi, F. Vincent, B. Volpe, R. Guidetti, N. Harris, A. Beck and G. Boero, *Scientific Reports*, 2017, **7**.
3. T. Stiernagle, *C. elegans*, 1999, **2**, 51-67.